\documentclass[12pt]{article}
\usepackage{graphicx}
\usepackage{color}

\def\Title#1{\begin{center} {\Large #1 } \end{center}}
\def\Author#1{\begin{center}{ \sc #1} \end{center}}
\def\Address#1{\begin{center}{ \it #1} \end{center}}

\newenvironment{Abstract}{\begin{quotation}  }{\end{quotation}}
\newenvironment{Presented}{\begin{quotation} \begin{center} 
             PRESENTED AT\end{center}\bigskip 
      \begin{center}\begin{large}}{\end{large}\end{center} \end{quotation}}
\def\Acknowledgements{\bigskip  \bigskip \begin{center} \begin{large}
             \bf ACKNOWLEDGEMENTS \end{large}\end{center}}

\def\beq{\begin{equation}}
\def\eeq#1{\label{#1}\end{equation}}
\def\eeqn{\end{equation}}

\def\beqa{\begin{eqnarray}}
\def\eeqa#1{\label{#1}\end{eqnarray}}
\def\eeqan{\end{eqnarray}}

\let\bar=\overbar

\def\Dslash{\not{\hbox{\kern-4pt $D$}}}
\def\dslash{\not{\hbox{\kern-2pt $\del$}}}

\def\msb{{\bar{\ssstyle M \kern -1pt S}}}

\begin{document}
\begin{titlepage}

\vfill
\Title{ The non-diffractive pp  cross-section and Survival Probabilities at LHC}
\vfill
\Author{Giulia Pancheri$^1$, A. Grau$^2$, Y. N. Srivastava$^3$, D. A. Fagundes$^4$ and O. Shekhovtsova$^5$}
\Address{$^1$INFN Frascati National Laboratories, Via Enrico Fermi 40, I00044 Frascati, Italy\\
$^2$ Departamento de Fisica Teorica y del Cosmos,
Universidad de Granada, 18071 Granada, Spain\\
$^3$
Physics Department, Northeastern University, Boston, Mass 02115, USA \\
$^4$ Departamento de Ci\^{e}ncias Exatas e Educa\c{c}\~{a}o, Universidade Federal de Santa Catarina -  Campus Blumenau, 89065-300, Blumenau, SC, Brazill\\
$^5$ NSC KIPT, Kharkov, 61108, Ukraine, 
and 
INP of PAS, Cracow, 31-234 Poland}
\vfill
\begin{Abstract}
We present an estimate of  survival probability  from an eikonal  mini-jet model implemented with a proposal for soft gluon resummation to all orders.  We compare it with experimental data for diffractive di-jet production from LHC experiments, CMS and ATLAS, both at LO and NLO order.
\end{Abstract}
\vfill
\begin{Presented}
EDS Blois 2017, Prague, \\ Czech Republic, June 26-30, 2017
\end{Presented}
\vfill
\end{titlepage}
\def\thefootnote{\fnsymbol{footnote}}
\setcounter{footnote}{0}

\section{Introduction}

 In this short note, we present an estimate for survival probability factors based on the  eikonal mini jet model implemented with soft gluon resummation in the infrared region, developed in \cite{Grau:1999,Godbole:2005}. We   call this    the Bloch-Nordsieck (BN) model, as its core feature is   the treatment  of  the infrared region through exponentiation of the divergent low frequency spectra  \cite{BN}.
\section{The BN model for pp total cross-section and its application to SP}

As  the center-of-mass energy available to  experimental physics increases, presently being  13 TeV at particle accelerators \cite{TOTEM,ATLAS,CMS} and $95^{+5}_{-8} TeV$ TeV from cosmic ray experiments \cite{Abbasi:2015},  estimates and models for the total $pp$ cross-sections (elastic and inelastic)  need to take into account the contribution from perturbative QCD. The proposal in \cite{Grau:1999,Godbole:2005} embeds a minij-jet contribution calculated from 
\begin{equation} 
\sigma_{mini-jet} (s)=\int_{p_{tmin}}^{\sqrt{s}/2} d p_t \int_{4
p_t^2/s}^1 d x_1 \int_{4 p_t^2/(x_1 s)}^1 d x_2 
\sum_{i,j,k,l}
f_{i|A}(x_1,p_t^2) f_{j|B}(x_2, p_t^2)
 \frac { d \hat{\sigma}_{ij}^{ kl}(\hat{s})} {d p_t}.
\label{eq:minijet}
\end{equation} 
into an eikonal formulation, where unitarity is respected and multiple parton-parton collisions can be taken into account: 
\begin{equation}
\sigma_{total}=2\int d^2{\bf b}[1-e^{-\chi_I(b,s)}] \label{eq:sigtot}
\end{equation}
In Eq.~(\ref{eq:minijet}), the parton densites are standard LO DGLAP evolved with $Q^2=p_t^2$,  the parton-parton cross-section is calculated from LO QCD and  the minimum parton momentum for which this equation can be used is $p_{tmin} \simeq (1\div 2)$ GeV. The impact parameter dependence in Eq.~(\ref{eq:sigtot}) is to be defined by the specific modeling, such as convolution of proton form factors or Fourier transform of soft gluon resummed contribution to single parton-parton collisions.
In the  BN model, as described in  our most recent Ref.\cite{Fagundes:2017},  we distinguish between {\it soft} and {\it hard} collisions. Correspondingly, we propose a {\bf b}-distribution from proton form factors for soft collisions parametrized with no pQCD input,  whereas for hard collisions, calculated through Eq.~(\ref{eq:minijet}),   the {\bf b}-distribution matches  
the expression obtained through  soft gluon resummation in  impact parameter space.  More precisely, the  proposal for the average number of collisions input to   calculation of the total   cross-section is given as
\begin{equation}
2\chi_I(b,s)={\bar n}(b,s)={\bar n}_{soft}(b,s) + {\bar n}_{hard}(b,s)=A_{FF}(b) \sigma_{soft}(s) + A_{BN}(b,s)\sigma_{mini-jet}(s)\label{eq:nbs}
\end{equation}
with 
\begin{equation}
A_{BN}(b,s)=N(s)
\ exp(-\int d^3 {\bar n}_g(k)[
1-e^{-ik_t\cdot {\bf b}}
]
)\label{eq:abn}
\end{equation}
In Eq.~(\ref{eq:nbs}) $\int d^2{\bf b} A_{BN/FF}(b,s)=1$, and, in Eq.~(\ref{eq:abn}), $d^3 {\bar n}_g(k)$ is the single soft gluon spectrum, which, in our model, is integrated from $k_t=0$ to a scale $q_{max}(s,p_{tmin};PDF)$, calculated in correspondence with the one gluon emission kinematics.  The integral over the single gluon spectrum can be extended into the $k_t=0$ limit if a suitable, integrable behavior for $d^3 {\bar n}_g(k)$ is possible. Given the presence of the infrared cancellation through  the factor between square brackets, this condition of integrabilitiy corresponds to an ansatz for the behavior of the strong coupling constant such that $\alpha_s(k_t\rightarrow 0) \propto k_t^{-2p}$, with $p<1$ \cite{Corsetti:96}. For a derivation of Eq.~(\ref{eq:abn}), we refer  the reader to \cite{Fagundes:2017} and references therein.

What we have presented in this section divides collisions into two simple groups, soft, and hard. The soft collisions are characterized as either constant or decreasing with energy, and corresponding to partons  described in impact parameter  space 
through  the proton form factor, i.e. $A_{FF}(b)$. The hard term rises with energy, and is obtained  from  collisions  between  two perturbative partons,   each one from  one of the  colliding protons, with b-distribution from soft gluon resummation.


\section{Survival probability in the BN model and comparison with other models}

\begin{figure}[htb]
  \includegraphics[height=2.6in]{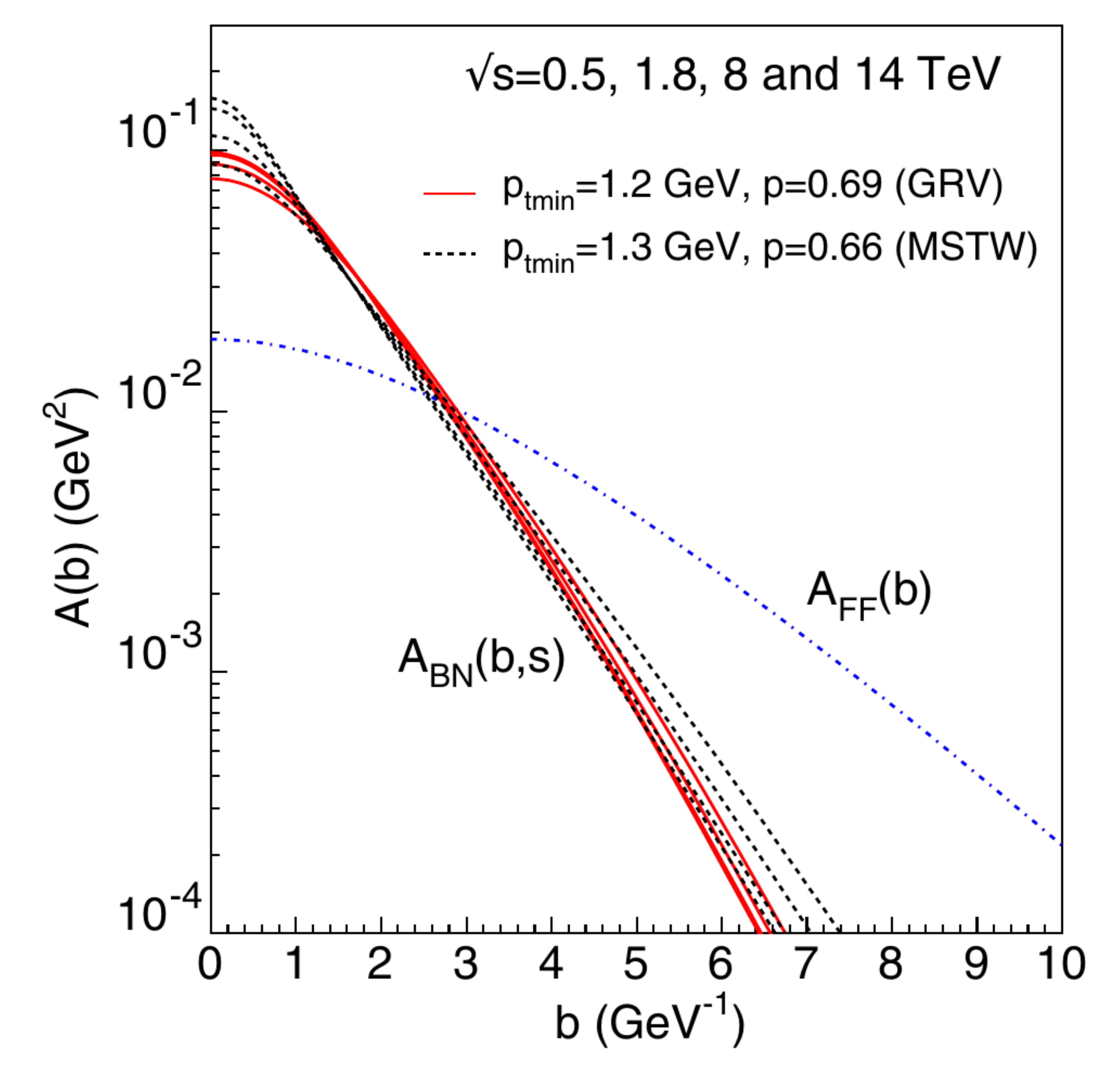}
 \includegraphics[height=2.6in]{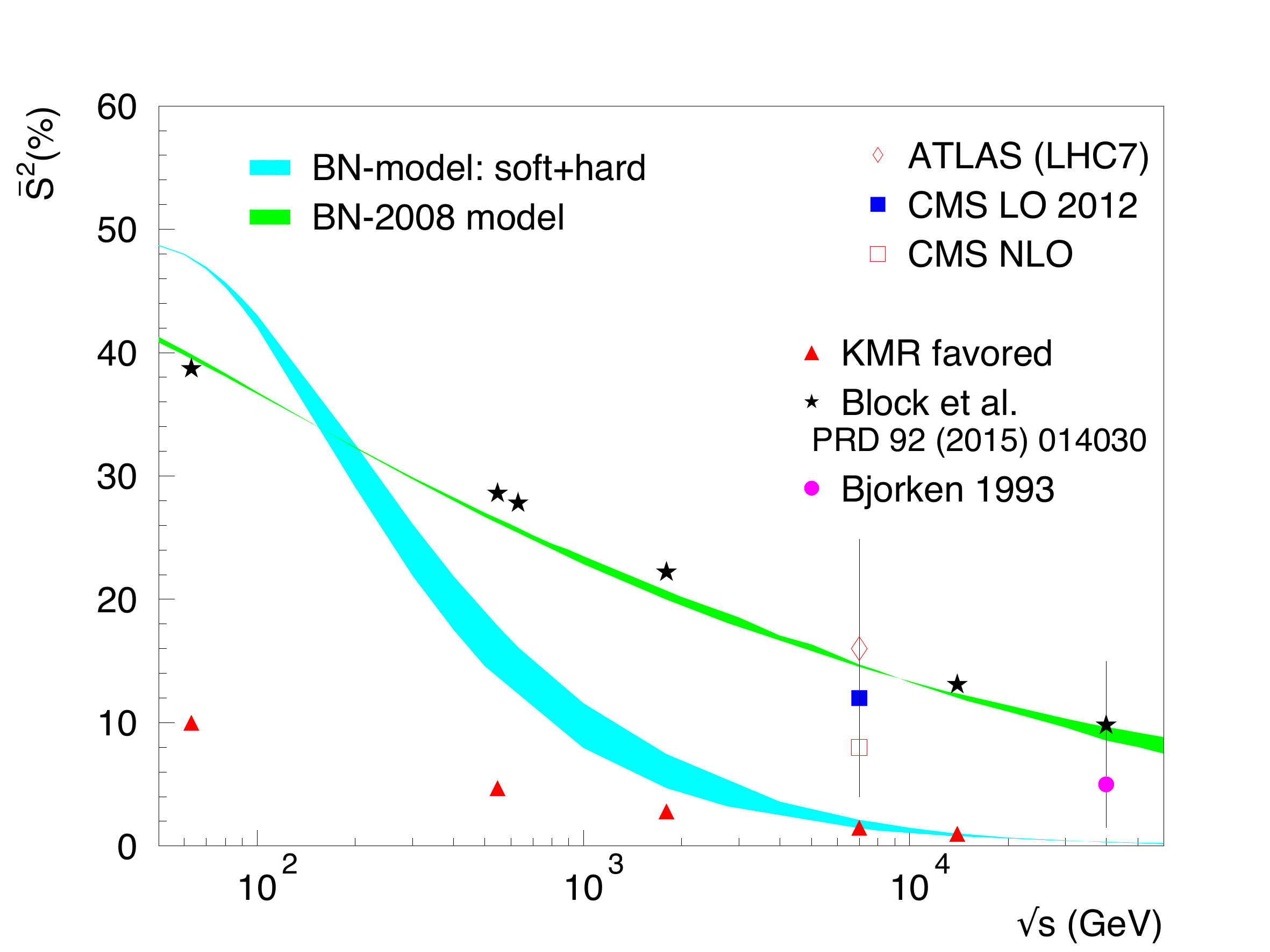}
 
 \caption{Left: soft and hard parton distribution in b-space. Right: our proposal for the Survival Probability compared to  estimates from  CMS \cite{CMS-SP} and ATLAS \cite{ATLAS-SP} extracted from diffractive jet production at LHC, and with three other models: Block's proposal in \cite{Block:2015},  a  2008 application of the BN-model from \cite{Godbole:2008}, estimates from Khoze, Martin and Ryskin (KMR) \cite{Khoze:2013}, and the Bjorken estimate \cite{Bjorken:1993}. For our present estimate,  two different PDFs parametrizations are used, GRV 
 and MSTW, spanning the indicated bands.
 Further details and references  can be found in Ref. \cite{Fagundes:2017}.}
  \label{fig:survprob}
 \end{figure}

The model outlined in the previous section can give a satisfactory description of the total cross-section, and of the inelastic non-diffractive cross-section \cite{Fagundes:2017}. As such, it  can be used to study survival probabilities for events with no hadronic activity outside the diffractive region. Following the original proposal  by Bjorken \cite{Bjorken:1993}, we shall calculate the survival probability using the expression
\begin{equation}
{\cal S}^2(s)\equiv<| S(b,s)|^2>=\int d^2{\bf b} A(b,s) P_{no-collisions}= 
\int d^2{\bf b} 
A(b,s)
e^{-{\bar n}(b,s)}\label{eq:survival} 
\end{equation} 
 To apply this expression to our mini jet model,  we   
 deem it  necessary to distinguish between the parton distributions 
 for {\it soft } and {\it hard} events, as 
 they have different energy and $b-$dependences. In particular, at  large distances, $A_{FF}(b)$ falls less rapidly 
than $A_{BN}(b,s)$; also,unlike $A_{BN}(b,s)$, the former  carries no energy dependence, as one can see in the left panel of Fig.~\ref{fig:survprob}.
 
In \cite{Fagundes:2017}, we have proposed to 
``weigh'' these different possibilities according to the different soft  and mini-jet type contributions as in  the following expression:
\begin{equation}
\bar{\mathcal{S}}^{2}_{total} (s)=\frac{\sigma_{soft}(s)}{\sigma_{soft}(s)+
\sigma_{mini-jet}(s)}
 <| S(b)|^2>_{soft}+\frac{\sigma_{mini-jet}(s)}{\sigma_{soft}(s)+
\sigma_{mini-jet}(s)}
 <| S(b)|^2>_{hard} 
\label{eq:survsum}
\end{equation}with
\begin{equation}
<| S(b)|^2>_{soft/hard}=\int d^2{\bf b} 
A_{FF/BN}(b,s)
e^{-{\bar n}_{soft/hard}(b,s)}
\end{equation}
In the context of mini-jet inspired models, the  possibility of a {\it mixed} expression has also been considered,  i.e.
\begin{equation}
{\cal S}^2(s)\equiv<| S(b,s)|^2>=\int d^2{\bf b} A_{soft}(b,s) P_{no-collisions}= 
\int d^2{\bf b} 
A_{soft}(b,s)
e^{-{\bar n}(b,s)}\label{eq:bn2008} 
\end{equation}
such as in \cite{Bjorken:1993,Block:2015,Godbole:2008}, with $ P_{no-collisions}$ obtained from a one-channel eikonal formulation and, in the case of \cite{Block:2015},  $A_{soft}(b,s)$ obtained from the quark-quark contribution, which is either  a constant or decreasing with energy. This expression would correspond to 
eliminating all  hadronic events for which the parton distribution is of the form-factor type -carrying no energy dependence. For the BN model, this corresponds  to 
eliminating all non-diffractive hadronic events, as we have mentioned.
In  the right panel of Fig.~\ref{fig:survprob}  we  show  our estimate from Eqs.~(\ref{eq:survsum}) and (\ref{eq:bn2008}), together with results from \cite{Bjorken:1993,Block:2015} and those from the model (favorite estimate) of   Khoze, Martin and Ryskin (KMR) \cite{Khoze:2013}. For comparison with KMR results as well as those from \cite{Gotsman:2015}, see Ref.\cite{Fagundes:2017}.
\subsection{Comments about applicability of the model}
The model outlined in the previous section can give a satisfactory description of the total cross-section, and of the inelastic non-diffractive cross-section, 
since
{its present one-channel formalism for the inelastic cross-section does not include  the diffractive component.  As a consequence, our estimates for the SP refer to experimental set-ups where only  non-diffractive events are excluded.}

 The pQCD description of the diffractive contribution  has 
different energy behavior than the one from parton-parton collisions.  Parametrizations of the contribution to  Single Diffractive events  obtained through multichannel versions of the Glauber model, indicate that  these contributions rise like $\log s$ \cite{Engel}. Experimentally, these events are also different, since they are correlated with one of the outgoing protons, peaking in a large rapidity region.

Using this model, we have addressed here the question of how different  impact parameter distributions affect estimates for  survival probability in eikonal mini-jets models.
 There are noticeable quantitative and qualitative differences between the two sets of theoretical predictions for SP. They reflect whether all hadronic activity beyond single diffraction has been removed or not.\\
So far, reported SP data exist only upto $7\ TeV$ showing little or no variations with energy. Data at\ $8\ \&\ 13\ TeV$ would also help resolve whether there is a sharp decline or not in SP for increasing energy.  



\Acknowledgements 

 We  acknowledge a  stimulating conversation with Valery Khoze about estimates of survival probabilities in different rapidity regions and for specific experimental cuts, such as for the CMS and ATLAS diffractive dijet data presented in the text. Y.S. thanks the Department of Physics and Geology at the University of Perugia for the continued
hospitality.
 A.G. acknowledges partial support by MINECO
 under grant number
FPA2016-78220-C3-3-P, and by Junta de Andaluc\'\i a (Grants FQM 101 and
FQM 6552). D.A.F.  acknowledges partial support from the project INCT-FNA Proc. No. 464898/2014-5.


\begin{thebibliography}{99}
\bibitem{Grau:1999} 
A. Grau, G. Pancheri, and Y. Srivastava, Phys.Rev. D60, 114020 (1999), arXiv:hep-
ph/9905228 [hep-ph].
\bibitem{Godbole:2005} R. M. Godbole, A. Grau, G. Pancheri, and Y. N. Srivastava, Phys. Rev. D72, 076001 (2005),
arXiv:hep-ph/0408355 [hep-ph].
\bibitem{BN} F. Bloch and A. Nordsieck, Phys. Rev. 52, 54 (1937).
\bibitem{TOTEM} G. Antchev et al. (TOTEM), Europhys. Lett. 101, 21004 (2013); G. Antchev et al. (TOTEM), Eur. Phys. J. C76, 661 (2016); G. Antchev et al. (TOTEM), Phys. Rev. Lett. 111, 012001 (2013).
\bibitem{ATLAS} G. Aad et al. (ATLAS Collaboration), Phys. Lett. B 754,
214 (2016).
\bibitem{CMS} S. Chatrchyan et al. (CMS Collaboration), Phys. Rev. D 87,
012006 (2013).
\bibitem{Abbasi:2015} R. U. Abbasi et al. (Telescope Array), Phys. Rev. D92, 032007 (2015), arXiv:1505.01860.
\bibitem{Fagundes:2017}  D.A. Fagundes, A. Grau, G. Pancheri, O. Shekhovtsova and Y.N. Srivastava, Phys.Rev. D96 (2017) no.5, 054010.
\bibitem{Corsetti:96} A. Corsetti, A. Grau, G. Pancheri and Y. N. Srivastava, Phys. Lett. B382, 282 (1996).
\bibitem{Bjorken:1993} J. D. Bjorken, Phys. Rev. D47, 101 (1993).
\bibitem{CMS-SP} S. Chatrchyan et al. (CMS), Phys. Rev. D87, 012006 (2013), arXiv:1209.1805 [hep-ex].
\bibitem{ATLAS-SP} G. Aad et al. (ATLAS), Phys. Lett. B754, 214 (2016), arXiv:1511.00502 [hep-ex].
\bibitem{Block:2015} M. M. Block, L. Durand, P. Ha, and F. Halzen, Phys. Rev. D92, 014030 (2015).
\bibitem{Godbole:2008} A. Achilli, R. Hegde, R. M. Godbole, A. Grau, G. Pancheri, and Y. Srivastava, Phys. Lett.
B659, 137 (2008).
\bibitem{Khoze:2013} V. A. Khoze, A. D. Martin, and M. G. Ryskin, Eur. Phys. J. C73, 2503 (2013),
\bibitem{Gotsman:2015} E. Gotsman, Levin E.  and U. Maor, Eur. Phys. J. C76, 177 (2016).
\bibitem{Engel}  R. Engel and R. Ulrich, Internal Pierre Auger Note GAP-2012 (March, 2012). 
   
\end{thebibliography}
\end{document}